\documentstyle[11pt,epsfig]{article}
\def\la{\mathrel{\mathchoice {\vcenter{\offinterlineskip\halign{\hfil
$\displaystyle##$\hfil\cr<\cr\sim\cr}}}
{\vcenter{\offinterlineskip\halign{\hfil$\textstyle##$\hfil\cr<\cr\sim\cr}}}
{\vcenter{\offinterlineskip\halign{\hfil$\scriptstyle##$\hfil\cr<\cr\sim\cr}}}
{\vcenter{\offinterlineskip\halign{\hfil$\scriptscriptstyle##$\hfil\cr<\cr\sim
\cr}}}}}

\newcommand{\be}{\begin{equation}}
\newcommand{\ee}{\end{equation}}
\newcommand{\bea}{\begin{eqnarray}}
\newcommand{\eea}{\end{eqnarray}}
\begin{document}
\thispagestyle{empty}
\begin{center}
{\Large\bf Quintom scalar field : 
\bf Varying dark energy equation of state obtained 
from recent SNe Ia, BAO and OHD data} \\
\end{center}
\vskip 1cm
\begin{center}
{\bf Debabrata Adak$^{\dagger}$\footnote{Email: debabrata.adak@saha.ac.in}, 
Abhijit Bandyopadhyay${^\ddagger}$\footnote{Email: abhi.vu@gmail.com} and  
Debasish Majumdar${^\dagger}$}\footnote{Email: debasish.majumdar@saha.ac.in}
\vskip 0.5cm
{\normalsize \it $\dagger$ Astroparticle Physics and Cosmology Division,} \\
{\normalsize \it Saha Institute of Nuclear Physics,} \\
{\normalsize \it 1/AF Bidhannagar, Kolkata 700064, India}\\
\vskip 0.2cm
{\normalsize \it $\ddagger$ Department of Physics} \\ 
{\normalsize \it Ramakrishna Mission Vivekananda University,} \\
{\normalsize \it Belur Math, Howrah 711202, India}\\
\end{center}

\begin{abstract}
From the analysis of Supernova Ia data alongwith Observational 
Hubble Data (OHD) and Baryon Acoustic
Oscillation (BAO) data, we attempt to find out the nature 
of a scalar potential that may be responsible for the Dark 
Energy of the universe. We demonstrate that in order to explain 
the varying dark energy equation of state ($\omega_X(z)$)
as obtained 
in a model independent way from the analyses of observational data
, we need to invoke 
a quintom scalar field 
having a ``quintessence" part for $\omega_X(z) > -1$ and a 
``phantom" part for $\omega_X(z) < -1$.
We consider a Gaussian type potential 
for these scalar fields and compare 
the dark energy equation of state derived from such potential with 
the one computed from the data analysis.    

\end{abstract}
\newpage

\section{Introduction}
The recent acceleration of the expansion rate of the universe 
that has been inferred
from the observation and luminosity distance measurements of different 
Supernova Ia (SNe Ia) \cite{sn} at different redshifts ($z$) is thought to be 
caused by the existence of a mysterious energy called dark energy. Different 
other observational evidence (such as WMAP \cite{wmap}, Cosmic Microwave 
Background (CMB) \cite{cmb}) etc.
suggest that this dark energy accounts for almost 73\% of the content of 
the universe. Dark energy is a hypothetical energy 
component with a negative pressure responsible for causing the 
universe to experience this accelerated expansion.
Although there are efforts to understand its nature 
and origin, dark energy still remains an enigma. 

An attempt to explain dark energy involvs considering a slowly varying
potential $V(\phi)$ of a scalar field $\phi$. 
Hence
for such considerations the knowledge of $\phi$ and $V(\phi)$ are necessary
in order to get an insight of dark energy and its equation of state.
In FRW cosmology, it can be shown that in cosmological constant scenario
, the dark energy equation of state remains 
constant at the value -1 all along the evolution history of 
the universe. However if dark energy varies with time or equivalently with
redshift $z$, a possible explanation may be given 
by a scalar field (quintessence field)
for dark energy equation of state $\omega(z) \geq -1$.
But one needs to 
invoke a phantom scalar field $\sigma$ in case $\omega(z)$ falls 
below  $-1$. 
  
In the present work,
we have considered five different SNe Ia data sets alongwith 
Observational Hubble Data (OHD) \cite{ohd}  and Baryon Acoustic Oscillation 
Data (BAO) \cite{bao}. The data sets are made available from 
Riess {\it et al} (2007)\cite{Riess07},
Wood-Vasey {\it et al} (2007)\cite{WoodVasey}, Davis {\it et al}\cite{Davis},
Kowalski {\it et al} (2008)\cite{kowalski}, Kessler {\it et al} (2009)
\cite{Kessler} and Amanullah {\it et al} (2010) (also known as
Union2 data)\cite {Amanullah}. The maximum span of redshift $z$ for these
compilations for distance modulus $\mu(z)$ (related to 
the luminosity distance $d_L(z)$) 
are in the range $0.001\leq z \leq 1.76$.
A parametric form for $d_L (z)$ is chosen whose parameters are
obtained from a $\chi^2$ minimization of various observational data 
mentioned above. The dark energy equation of state can be derived 
analytically using an analytic form of $d_L (z)$, such as the present 
parametric form. With the fitted values of the parameters the 
dark energy equation of state $\omega_X(z)$ can therefore be 
calculated. 
On the otherhand, since $d_L (z)$ is related to the Hubble parameter $H(z)$
by an analytical expression one can use the same parametric form
for $d_L (z)$ (with parameters from $\chi^2$ analyses) and evaluate 
both the quintessence 
scalar field $\phi$ and the phantom field $\sigma$ and their variations 
with the redshift $z$ (see later). With chosen form
of potentials for both $\phi$ and $\sigma$, one can then compute the 
dark energy equation of state ($\omega(z)$) from the $\phi$ and $\sigma$
values obtained from the analysis. In this work, we have taken the 
form of this potential as $V = V(\phi) + V(\sigma)$, where 
$V(\phi)$ and $V(\sigma)$ are chosen to be 
$V(\phi)=V_0\exp[\lambda_1 (1+(\phi(z)-\phi_0)^2/M^2_{pl})]$ and
$V(\sigma)=V_0\exp[\lambda_2 (1+(\sigma(z)-\sigma_0)^2/M^2_{pl})]$,
where $\phi_0$($\sigma_0$) is the value of $\phi$($\sigma$) at present epoch
and $M_{pl}$ is the Planck mass.
We refer the dark energy equation 
of state computed from the scalar fields with this potential, as $\omega(z)$. 
The $\omega(z)$ thus obtained is 
then compared with the $\omega_X(z)$ computed directly from the analysis 
of $\mu_L (z)$ data with the assumption that the present universe is
spatially flat and contains only matter and dark energy.

The paper is organised as follows. In Sect. 2 we discuss the dark energy
equation of state as derived from FRW cosmology. In Sect. 3 we propose a
quintom scalar formalism of dark energy. The calculational procedures are
given in Sect. 4. Finally in Sect. 5 we furnish our calculational results
and discussions.

\section{Brief overview of model independent reconstruction of dark energy
equation of state}

For a homogeneous and isotropic universe 
the Friedmann-Robertson-Walker metric 
is given by
\bea
ds^2&=& - dt^2+a^2(t)\left[\frac{dr^2}{1-Kr^2}+r^2(d\theta^2+sin^2\theta d\phi^2)\right] \,\, .
\label{frwmet1}
\eea 
where $a(t)$ is the scale factor and $K$ is the curvature parameter.
Considering the universe to be a perfect fluid characterized by energy 
density $\rho$ and pressure $p$, the Einstein's equation alongwith 
Eq. (\ref{frwmet1}) leads to two independent equations (for spatially
flat universe),
\bea
H^2(t) &=& \frac{8 \pi G}{3} \rho(t) \label{hub2}\\
\dot H(t) &=& - 4 \pi G (p+\rho) \label{dothub}
\eea
where $H(t)=\frac{1}{a(t)}\frac{d}{dt}(a(t))$ is the Hubble parameter
that denotes the expansion rate of universe.

From Eq. (\ref{hub2}) and Eq. (\ref{dothub}) one gets 
\bea
\frac{\ddot{a}(t)}{a(t)} = - \frac{4 \pi G}{3}(\rho+3p) \, .
\label{acce}
\eea
Assuming the universe contains only matter and dark energy, Eq. (\ref{hub2}) 
leads to the result
\bea
\frac{H^2(z)}{H_0^2}=\Omega_m^0(1+z)^3+\Omega_X^0\exp\left(\int_0^{z} 
3(1+\omega_X(z))\frac{dz}{1+z}\right)\,\, ,
\label{hub3}
\eea
where the symbol $\Omega$ represents energy density or matter 
density normalised 
to the critical density $\rho_c$ of the universe and thus $\Omega_m^0$  
($\Omega_X^0$) represents the matter (dark energy) density parameter
at present epoch. As mentioned earlier in this section that universe
contains only matter and dark energy in the redshift range considered here,
we have $\Omega_m(z) + \Omega_X(z) = 1$.
From Eq. (\ref{hub3}) the equation of state of dark 
energy $\omega_X(z)$ can be derived as, 
\bea
\omega_X(z) &=& - 1 + \left [\frac{\frac{2}{3}\frac{(1+z)}{H(z)}\frac{dH(z)}
{dz} - 
\Omega_m^0\frac{(1+z)^3}{H^2(z)}H_0^2}{1-\Omega_m^0\frac{(1+z)^3}{H^2(z)}H_0^2}
\right ]
\label{omegX1}
\eea

The SNe Ia observational data are tabulated as a distance modulus
$\mu(z)$. This is related
to luminosity distance $d_L(z)$ by
\bea 
\mu(z)&=&5 \log_{10}(d_L/Mpc) + 25
\label{mu}
\eea
 for a redshift value $z$ at which the 
observation was made.
The observed distance modulus $\mu_{\rm obs}(z)$ is related to the apparent 
magnitude $m_{\rm obs}(z)$ of a SNe Ia and absolute magnitude $M$ through the
equation $\mu_{\rm obs}(z)=m_{\rm obs}(z) - M$. Also 
for spatially flat universe Hubble parameter and the luminosity distance
are related through
\bea
H(z)&=&c\left[\frac{d}{dz}\left(\frac{d_L(z)}{(1+z)}\right)\right]^{-1} \,\, .
\label{hub1}
\eea
As stated in Sect. 1 a parametric form of $d_L(z)$ is considered and the 
parameters are obtained from a $\chi^2$ fit of the
observational data. 
The dark energy equation of state $\omega_X(z)$ can then be reconstructed 
using the Eqns (\ref {frwmet1} - \ref {hub3}) and Eq. (\ref {omegX1}).    
We also mention here that a combined $\chi^2$ fit is made with SNe Ia data, 
OHD and BAO data.

\section{Dark energy equation of state in a scalar field theory framework}
The acceleration equation (Eq. (\ref{acce})) suggests that the universe 
containing a perfect 
fluid can undergo accelerated expansion only if the pressure of the fluid
is negative. A suitably chosen real scalar field can produce negative 
pressure in the FRW spacetime and may also lead to time varying equation of
state of the perfect fluid depending on the potential of the field. Therefore 
such a real scalar field can be a possible candidate for time varying dark
energy. The scalar fields naturally arise in particle physics theories 
including 
string theory. There exists in literature a wide variety of 
 scalar field 
dark energy models. In this work we consider a quintom scalar 
field model where the two scalar fields namely $\phi(z)$ and $\sigma(z)$ are 
responsible for dark energy and the variation of its equation of state
with redshift $z$. Each of the fields has specified form of potential
and the two kinds of fields are not coupled to each other.
We like to show that this simple model can explain the varying dark energy 
equation of state $\omega_X(z)$ obtained by using the combined $\chi^2$ 
analyses results of SNe data, OHD data, BAO data in Eq. (\ref{omegX1}).

\subsection{Dark energy and quintessence scalar field}
The action $S$ for Quintessence scalar field is given by (with $V(\phi)$, 
the potential of the field $\phi$),
\bea
S&=& S_m + \int d^4x\sqrt{-g}\left[\frac{1}{2}g^{\mu\nu}(\partial_\mu\phi)
\partial_\nu\phi) + V(\phi)\right] \,\, ,
\label{actionqu}
\eea
where $S_m$ is the matter action. 
The equation of motion of the spatially homogeneous quintessence field 
is given by
\bea
\ddot\phi(t)+3H(t)\dot\phi(t)+\frac{dV(\phi)}{d\phi}&=& 0\,\,\,\,.
\label{eomqu}
\eea
The Eq. (\ref{eomqu}) will lead to
\bea
\dot\phi^2(t)=-\frac{1}{8\pi G H(t)}\frac{d}{dt}(H_{\phi}^2)\,\,\,\,,
\label{kequ}
\eea
where $H_{\phi}^2=\frac{8\pi G}{3}\left(\frac{1}{2}\dot\phi^2(t)+V(\phi)
\right)$ and $H^2=\frac{8\pi G}{3}\left(\frac{1}{2}\dot\phi^2(t)+V(\phi)
+\rho_m\right)$.
The Energy momentum tensor for quintessence scalar field can be written as
\bea
T^\mu_\nu&=&g^{\mu\alpha}\partial_\alpha\phi\partial_\nu\phi-\delta^\mu_\nu
\left(\frac{1}{2}g^{\alpha\beta}\partial_\alpha \phi\partial_\beta \phi+
V(\phi)\right)
\label{enmotenqu}
\eea

From Eq. (\ref{enmotenqu}) we have,
\bea
\rho_{\phi}&=&-T^0_0=\frac{1}{2}\dot\phi^2+V(\phi)\\
\label{rhoqu}
p_{\phi}&=&T^i_i=\frac{1}{2}\dot\phi^2-V(\phi)
\label{pqu}
\eea
\bea
\omega_{\phi}(t)=\frac{p_\phi}{\rho_\phi}=\frac{\frac{1}{2}\dot\phi^2-
V(\phi)}{\frac{1}{2}\dot\phi^2+V(\phi)}
\label{omqu}
\eea
where $\rho_\phi$ and $p_\phi$ are respectively the pressure and density
due to the scalar field $\phi$. 
It is evident from Eq. (\ref{omqu}) that $\omega_\phi$ can vary 
between -1 and +1.

\subsection{Dark energy and phantom scalar field}
The action for Phantom scalar field $\sigma(t)$ is given by (with $V(\sigma)$,
the potential of the field $\sigma$),
\bea
S&=& S_m + \int d^4x\sqrt{-g}\left[ - \frac{1}{2}g^{\mu\nu}(\partial_\mu\sigma)
\partial_\nu\sigma) + V(\sigma)\right] \,\, .
\label{actionph}
\eea
where $S_m$ is the matter action.
The equation of motion of the spatially homogeneous phantom field
can be written as
\bea
\ddot\sigma(t)+3H(t)\dot\sigma(t)-\frac{dV(\sigma)}{d\sigma}&=& 0\,\,.
\label{eomph}
\eea
The above equation leads to
\bea
\dot\sigma^2(t)=\frac{1}{8\pi G H(t)}\frac{d}{dt}(H_{\sigma}^2)\,\,\,\,,
\label{keph}
\eea
where $H_{\sigma}^2=\frac{8\pi G}{3}\left(\frac{1}{2}\dot\sigma^2(t)+V(\sigma)
\right)$ and $H^2 = \frac{8\pi G}{3}\left(\frac{1}{2}\dot\sigma^2(t)+V(\sigma)
+\rho_m\right)$.
The Energy momentum tensor for phantom scalar field is given by
\bea
T^\mu_\nu&=&-g^{\mu\alpha}\partial_\alpha\sigma\partial_\nu\sigma-
\delta^\mu_\nu\left(\frac{1}{2}g^{\alpha\beta}\partial_\alpha \sigma
\partial_\beta \sigma+V(\sigma)\right)\,\, .
\label{enmotenph}
\eea
From Eq. (\ref{enmotenph}) we have,
\bea
\rho_{\sigma}&=&-T^0_0=-\frac{1}{2}\dot\sigma^2+V(\sigma)\\
\label{rhoph}
p_{\sigma}&=&T^i_i=-\frac{1}{2}\dot\sigma^2-V(\sigma)
\label{pph}
\eea
\bea
\omega_{\sigma}(t)=\frac{p_\sigma}{\rho_\sigma}=
\frac{-\frac{1}{2}\dot\sigma^2-V(\sigma)}{-\frac{1}{2}\dot\sigma^2+V(\sigma)}
\label{omph}
\eea
The above equation shows that $\omega_\sigma$ can vary between -1 and -$\infty$.

\subsection{Quintom scalar field motivated dark energy equation of state}
We consider here a model \cite{Guo} which contains a negative kinetic 
scalar field $\sigma$
and a normal scalar field $\phi$ with a general potential, described by
\bea
S = S_m + \int d^4x \sqrt {-g} \left [ \frac{R}{16 \pi G} - \frac{1}{2}
g^{\mu\nu}\partial_\mu\sigma\partial_\nu
\sigma + \frac{1}{2}g^{\mu\nu}\partial_\mu
\phi\partial_\nu\phi + V(\sigma,\phi) \right]\,\,\,\, ,
\eea
where $S_m = \int d^4x {\cal L}_m$ and ${\cal L}_m$ represents the 
Lagrangian density of 
matter fields.
Assuming the fields to be homogeneous, in a spatially flat FRW cosmological
model the 
equation of motion of the fields and the matter density are given by
\bea
\ddot\phi(t)+3H(t)\dot\phi(t)+\frac{dV(\phi)}{d\phi}&=& 0\\
\ddot\sigma(t)+3H(t)\dot\sigma(t)-\frac{dV(\sigma)}{d\sigma}&=& 0\\
\dot \rho_\gamma + 3H(t)(\rho_\gamma + p_\gamma) &=& 0 \,\,,
\eea
where the $\rho_\gamma$ is the density of fluid (matter) with a barotropic
equation of state $ p_\gamma = (\gamma - 1)\rho_\gamma$, $\gamma$ being 
a constant with $0 \le \gamma \leq 2$. For radiation, $\gamma = 4/3$ and for 
dust $\gamma = 1$.  
The corresponding effective pressure and energy densities will be
\bea
p &=& -\frac{1}{2}\dot{\sigma}^2 + \frac{1}{2}\dot{\phi}^2 - V(\sigma,\phi) \\
\rho &=& -\frac{1}{2}\dot{\sigma}^2 + \frac{1}{2}\dot{\phi}^2 + V(\sigma,\phi)
\eea
In this formalism the Friedmann equation becomes 
\bea
H^2&=&\frac{8 \pi G}{3}\left(- \frac{1}{2}\dot\sigma ^2 + V(\sigma) + 
\frac{1}{2}\dot\phi ^2 + V(\phi) + \rho_\gamma \right) \nonumber \\
&=& H_\sigma^2 + H_\phi^2 + \frac{8 \pi G}{3} \rho_\gamma
\label{hubquintom}
\eea
and the equation of state parameter $\omega(z)$ as obtained
from the present quintom scalar field formalism will then be 
written as
\bea
\omega &=& \frac{ - \dot \sigma ^2 + \dot \phi ^2 - 2V(\sigma,\phi)}
{- \dot \sigma ^2 + \dot \phi ^2 + 2V(\sigma,\phi)} \, .
\label{quinomega}
\eea
If the two scalar fields ($\sigma$ and $\phi$) are not directly coupled
to each other then $V(\sigma,\phi)$ can be written as
$V(\sigma,\phi) = V(\sigma) + V(\phi) $. With this assumption one can
readily see that for $\dot \phi \geq \dot \sigma$, Eq. (\ref{quinomega})
gives to $\omega \geq - 1$ implying the scenario for quintessence
scalar field and for $\dot \phi < \dot \sigma$,  $\omega < - 1$  signifying
the phantom scalar field model.

\section{Calculational Procedure}
The purpose of this work is to show the viability of the 
present formalism of quintom
scalar field model in explaining the dark energy and the variation of
its equation of state with $z$.
In order to obtain the dark energy equation of state from observational data,
 a parametric form for $d_L(z)$ \cite{paddy, taup} is first considered 
whose parameters are fixed by the $\chi^2$ minimization of the combined
data (SNe Ia, OHD, BAO). 
The matter density at the present epoch $\Omega_m^0$ is also made a
parameter in this $\chi^2$ analyses and its value is obtained from the
same $\chi^2$ minimization.
The form of $d_L(z)$ can now be used to 
calculate Hubble parameter $H(z)$ using Eq. (\ref{hub1}). Eq. (\ref{omegX1})
now readily gives the dark energy equation of state $\omega_X(z)$.

The parametric form of $d_L(z)$ should respect
the conditions that $d_L(z)=0$ at $z=0$ and $d_L(z) \propto z$ for
large $z$.
SNe Ia data are available for $z \la 1.76$ from various supernova 
observations. A parametrised form of $d_L(z)$ can be written
as \cite{paddy, taup}
\begin{eqnarray}
d_L(z) &=& \frac{c}{H_0} \left[\frac{z(1+az)}{1+bz}\right]
\label{dlzparam}
\end{eqnarray}
The parameters $a$, $b$ and $\Omega^0_m$ are obtained by minimising 
a suitably defined $\chi^2$. In the present case the $\chi^2$ is defined as
\cite{taup}, 
\bea
\chi^2_{\rm tot}=\chi^2_{\rm SN} + \chi^2_{\rm BAO} + \chi^2_{\rm OHD}
\label{chitot}
\eea
where $\chi^2_{\rm SN}$ corresponds to the SNe Ia data, $\chi^2_{\rm BAO}$
defines the  $\chi^2$ for Baryon acoustic oscillation data and 
$\chi^2_{\rm OHD}$ represents the $\chi^2$ for Observational Hubble Data 
(OHD). 

$\chi^2_{\rm SN}$ \cite{snchisq} is defined as 
\bea
\chi^2_{\rm SN}(a,b,M^{\prime})=\sum_{i=1}^N \frac{(\mu_{\rm obs}(z) - 
\mu_{\rm th}(a,b,z))^2}{\sigma_i^2}\,\, .\nonumber
\eea
In the above, the theoretical distance modulus $\mu_{\rm th}(a,b,z)$ can 
be obtain from Eq. (\ref{mu}).  
The above equation can be written in the form \cite{snchisq}
\bea
\chi^2_{\rm SN}(a,b,M^{\prime}=B/C)=A-\frac{B^2}{C}\,\,\,\,.
\label{chi2}
\eea
where
\bea
A&=&\sum_{z_i}\frac{(5\log_{10}(D_L(a,b,z_i))-m_{obs}(z_i))^2
}{\sigma_i^2}\,\,\,\,,\\
\label{A}
B&=&\sum_{z_i}\frac{(5\log_{10}(D_L(a,b,z_i))-m_{obs}(z_i))
}{\sigma_i^2}\,\,\,\,,\\
\label{B}
C&=&\sum_{z_i}\frac{1}{\sigma_i^2}\,\,\,\,.
\label{C}
\eea
In the above $D_L(z)=H_0d_L(z)/c$
and $\sigma_i$ are the errors. 

For the case of BAO data $\chi^2$ is defined as
\bea
\chi^2_{\rm BAO}(\Omega_m^0,a,b) &=& \frac{[A(\Omega_m^0,a,b) - 0.469]^2}
{0.017^2}
\label{chibao}
\eea
where
\bea
A(\Omega_m^0,a,b) &=& \frac{\sqrt{\Omega_m^0}}{[H(z_1)/H_0]^{1/3}}
\left [ \frac{1}{z_1} \int_0^{z_1} \frac{dz}{H(z)/H_0}\right ]^{2/3}
\eea
The experimental value of 
$A = 0.469\pm 0.017$ \cite{bao} in the above is obtained from the 
Sloan Digital Sky Survey (SDSS) \cite{bao} data at $z=0.35$. 

For OHD data, $\chi^2$ is defined as
\bea
\chi^2_{\rm OHD} &=& \sum_{i=1}^{15} \left [
\frac {H_{\rm th}(a,b;z_i) - H_{\rm obs}(z_i)} {\sigma_i} \right ]^2\,\, .
\eea
In the above, $\sigma_i$ denote the errors and $H_{\rm th}(a,b;z)$ is 
obtained as (using Eqs. \ref{hub1} and \ref{dlzparam})
\bea
H_{\rm th}(a,b;z)&=&H_0 \frac {[(1 + az)(1 + bz)]^2}
{(1 + 2az + (ab + a - b)z^2)} \,\, .
\label{hth}
\eea 
The parameters $a, b$ (and $\Omega_m^0$) are obtained by minimizing 
the $\chi^2$ defined above with SNe Ia, OHD and BAO data together. 
As described earlier in Sect. 1 we consider five data sets 
of SNe Ia. Thus we have five combined data sets of
(SNe Ia + OHD + BAO) and from the $\chi^2$ minimization 
five sets of parameters ($a$, $b$, 
$\Omega_m^0$) are obtained \cite{taup}. 
Needless to mention that for all the five combined data sets OHD and BAO
data remain the same.

\subsection{Variation of $\omega(z)$ with scalar fields}
The dark energy equation of state obtained 
from the present quintom scalar field model is denoted as $\omega(z)$.
The variation of $\omega(z)$
with scalar fields can be obtained 
using Eq. (\ref{quinomega}) with a chosen form for the scalar 
potentials. In the present work the form of the potentials for the fields 
$\phi$ and $\sigma$ are 
taken as $$V(\phi) = \exp[\lambda_1(1 + (\phi/M_{pl})^2)]$$ 
and $$V(\sigma)=\exp[\lambda_2(1 + (\sigma/M_{pl})^2)]$$ 
where $M_{pl}(=1/8\pi G)$ is the Planck mass and $\lambda_1$ 
and $\lambda_2$ are two parameters.

From Eq. (\ref{kequ}), we obtain for quintessence scalar field
\bea
(\phi(z) - \phi_0)/M_{pl}&=& - \int_0^z \frac{dz}{(1 + z)H(z)}\sqrt{
(1 + z)\frac{d}{dz}(H_{\phi}^2)}\,\,\,\,,
\label{expqu}
\eea
and from Eq. (\ref{keph}) we get for phantom field as
\bea
(\sigma(z) - \sigma_0)/M_{pl} &=& - \int_0^z \frac{dz}{(1 + z)H(z)}\sqrt{-(1 + z)
\frac{d}{dz}
(H_{\sigma}^2)}\,\,\,\,,
\label{expph}
\eea
where $\phi_0$ and $\sigma_0$ 
are values of the fields at present epoch. With the parameters 
$a$ and $b$ obtained from the $\chi^2$ fit, 
the Hubble parameter
$H(z)$ appearing in Eqs. (\ref{expqu}, \ref{expph}) can be 
computed from Eq. (\ref{hth}).  

As the quantities $H_\phi$ and $H_\sigma$ cannot directly be extracted 
from data, it is difficult to obtain $\phi(z)$ and $\sigma(z)$ form 
Eqs. (\ref{expqu}, \ref{expph}). It is also not possible to choose the 
boundary conditions of both the fields from observations. In order 
to circumvent this problem we propose the following formalism.  
For a varying dark energy density $\omega(z)$ with redshift $z$, 
when $\omega(z) > -1$, the quintessence scalar field $\phi(z) >> \sigma(z)$ and 
in the case when $\omega(z) < -1$, the phantom scalar field 
$\sigma(z) >> \phi(z)$. In the former case therefore ($\omega(z) > -1$),
 $\sigma(z)$ can be
taken to be zero for all practical purposes and similarly for the 
latter case ($\omega(z) < -1$), $\phi(z)$ can be taken to be zero. 
Thus, defining 
$z_c$ as the value of redshift $z$ at which $\omega(z) = -1$,
we can write, using this formalism, that in the redshift limit when
$z > z_c$, $\sigma(z) =0$ and when $z < z_c (z \geq 0$), $\phi(z) =0$.
At $z = z_c$, however, $\phi(z) = \sigma(z)$ in order that the 
continuity of the variation of $\omega(z)$ with $z$ is not affected.

Under this formalism, for $z > z_c$, 
the quintessence scalar field can be derived from Eq. (\ref{kequ}) as,
\bea
(\phi(z) - \phi_c)/M_{pl}&=& - \int_{z_c}^z \frac{dz}{(1 + z)H(z)}\sqrt{
(1 + z)\frac{d}{dz}(H_{\phi}^2)}\,\,\,\,,
\label{expqu1}
\eea
where $H^2=H^2_{\phi}+\frac{8\pi G}{3}\rho_\gamma$ 
(neglecting $H^2_{\sigma}$ in Eq. (\ref{hubquintom})).
Also, for the case $z < z_c, z \geq 0$, we get
from Eq. (\ref{keph}) the phantom field as 
\bea
(\sigma(z) - \sigma_0)/M_{pl}&=& - \int_0^z \frac{dz}{(1 + z)H(z)}\sqrt{-(1 + z)
\frac{d}{dz}
(H_{\sigma}^2)}\,\,\,\,,
\label{expph1}
\eea
where $H^2=H^2_{\sigma}+\frac{8\pi G}{3}\rho_\gamma$ 
(neglecting $H^2_{\phi}$ in Eq. (\ref{hubquintom})).
In the present work we are considering the redshift limit 
$0 < z < 1.76$ $-$ the present observational reach of SNe Ia. In this 
epoch the universe is supposed to be dominated by dark matter and 
dark energy and therefore $\rho_\gamma$ in Eq. (\ref{hubquintom}) 
denotes the matter
density $\rho_m$. Hence the matter density can be written
in terms of redshift $z$ as 
$\frac{8\pi G}{3}\rho_m = H_0^2 \Omega_m^0(1+z)^3$,
where $H_0$ and $\Omega_m^0$ are the Hubble parameter and matter density 
parameter in the present epoch respectively.

Now as mentioned earlier in the beginning of this section, 
$\omega_{\sigma,\phi}(z)$ can be calculated from Eq. (\ref{quinomega})
with potentials $\exp[\lambda_1(1 + ((\phi-\phi_c)/M_{pl})^2)]$ and 
$\exp[\lambda_2(1 + ((\sigma-\sigma_0)/M_{pl})^2)]$ for different choices of 
$\lambda_1$ and $\lambda_2$.

\section{Results and Discussions}
\begin{figure}
\includegraphics[height=8cm,width=8cm,angle=-90]{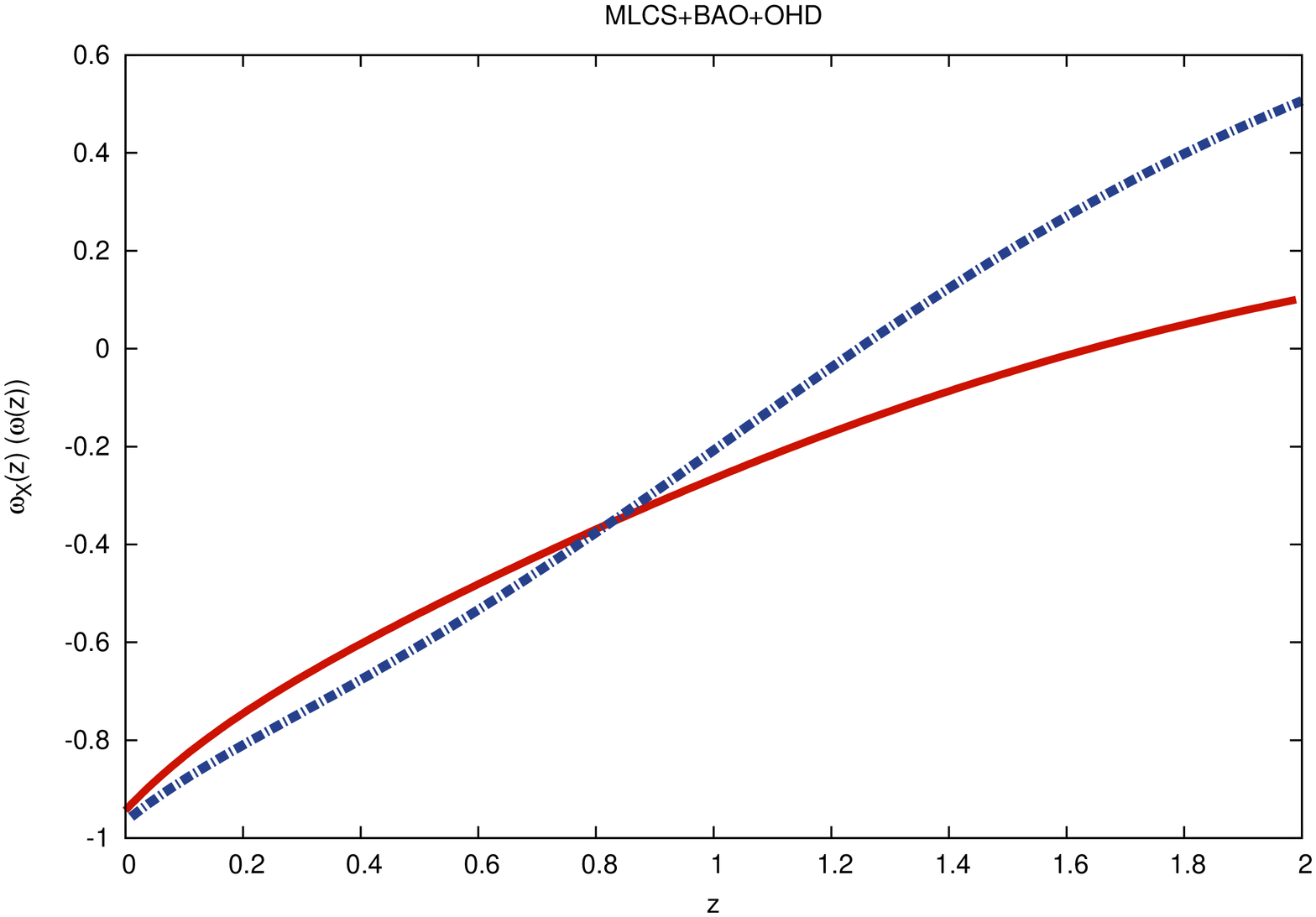}
\includegraphics[height=8cm,width=8cm,angle=-90]{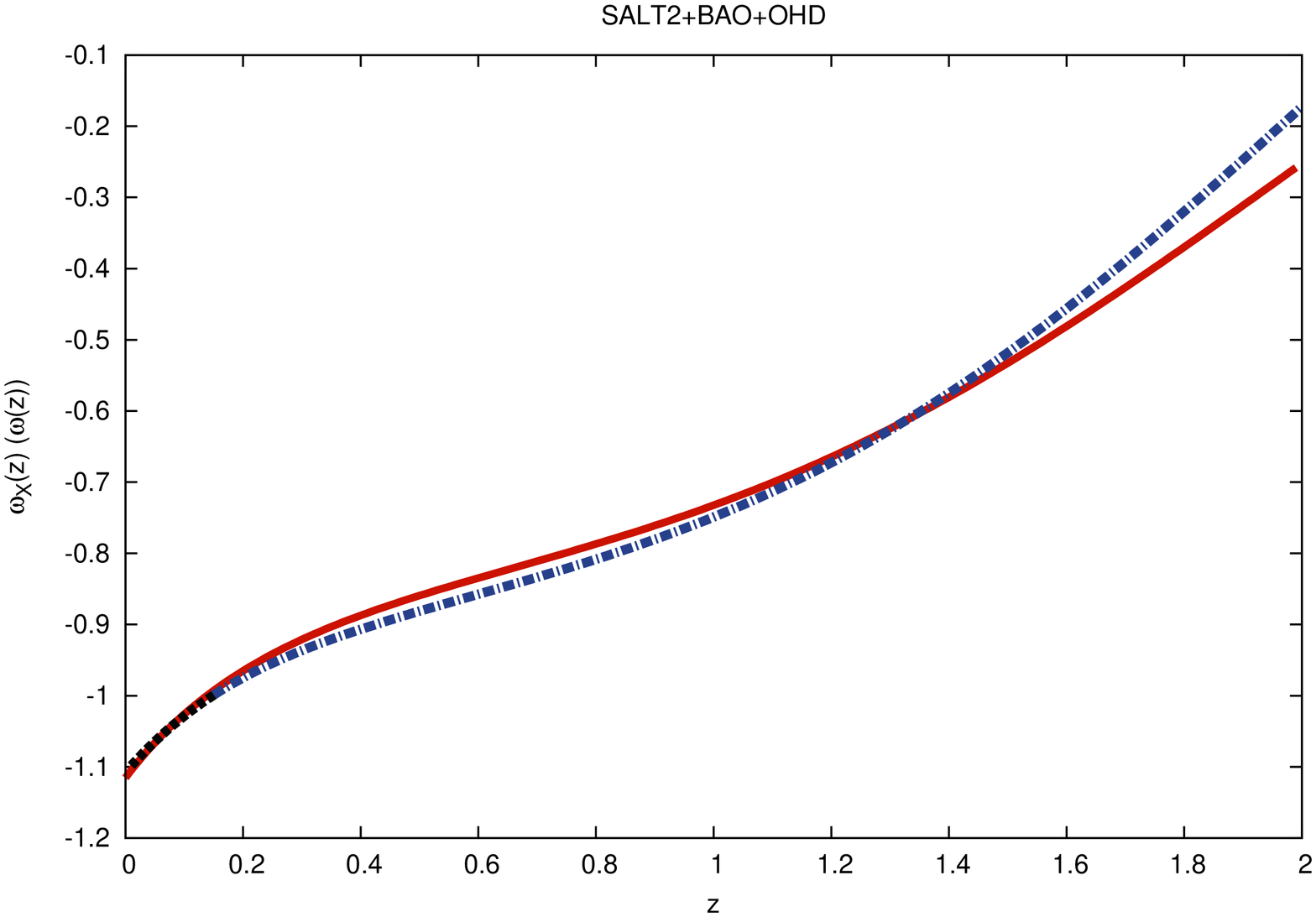}
\includegraphics[height=8cm,width=8cm,angle=-90]{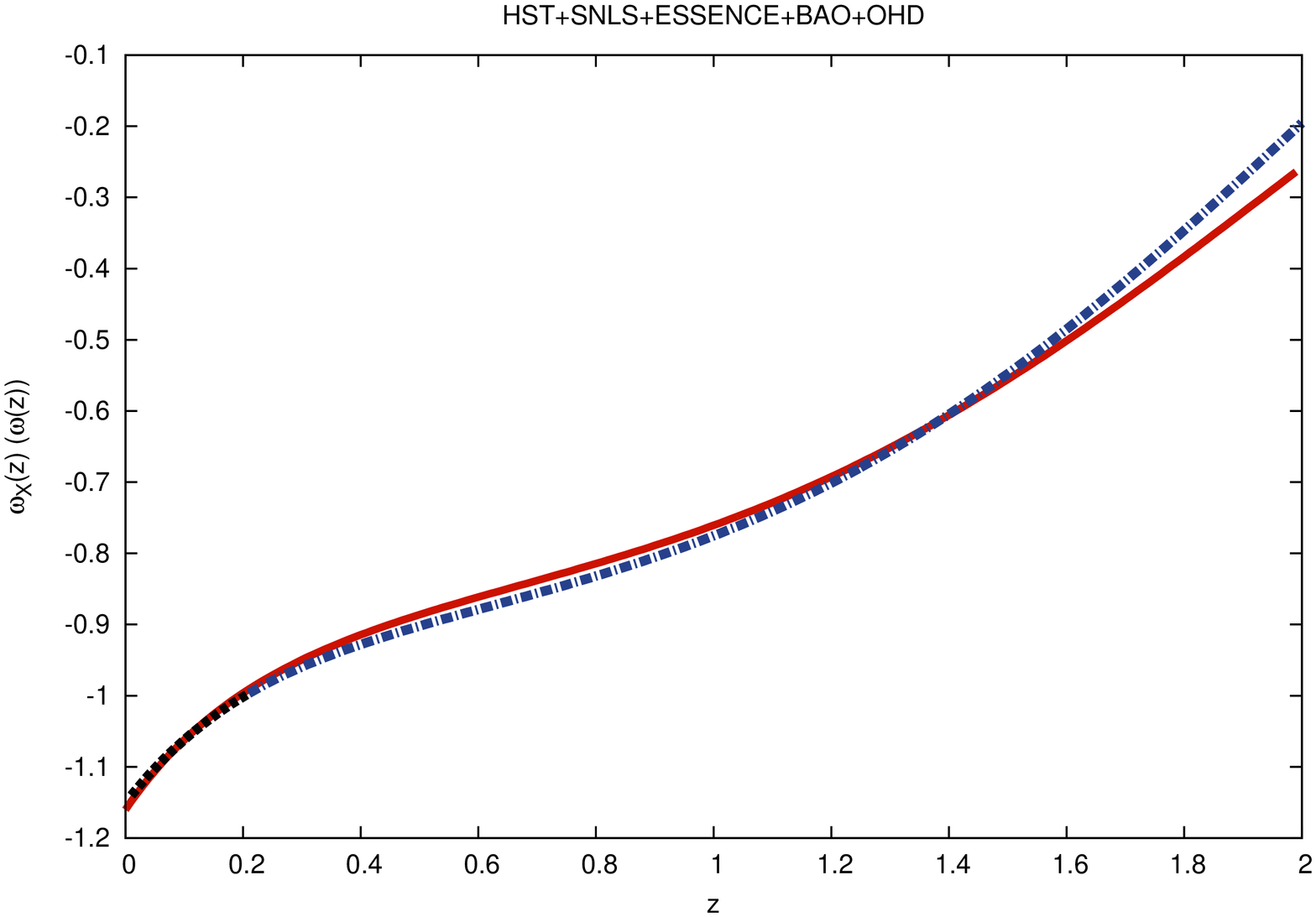}
\includegraphics[height=8cm,width=8cm,angle=-90]{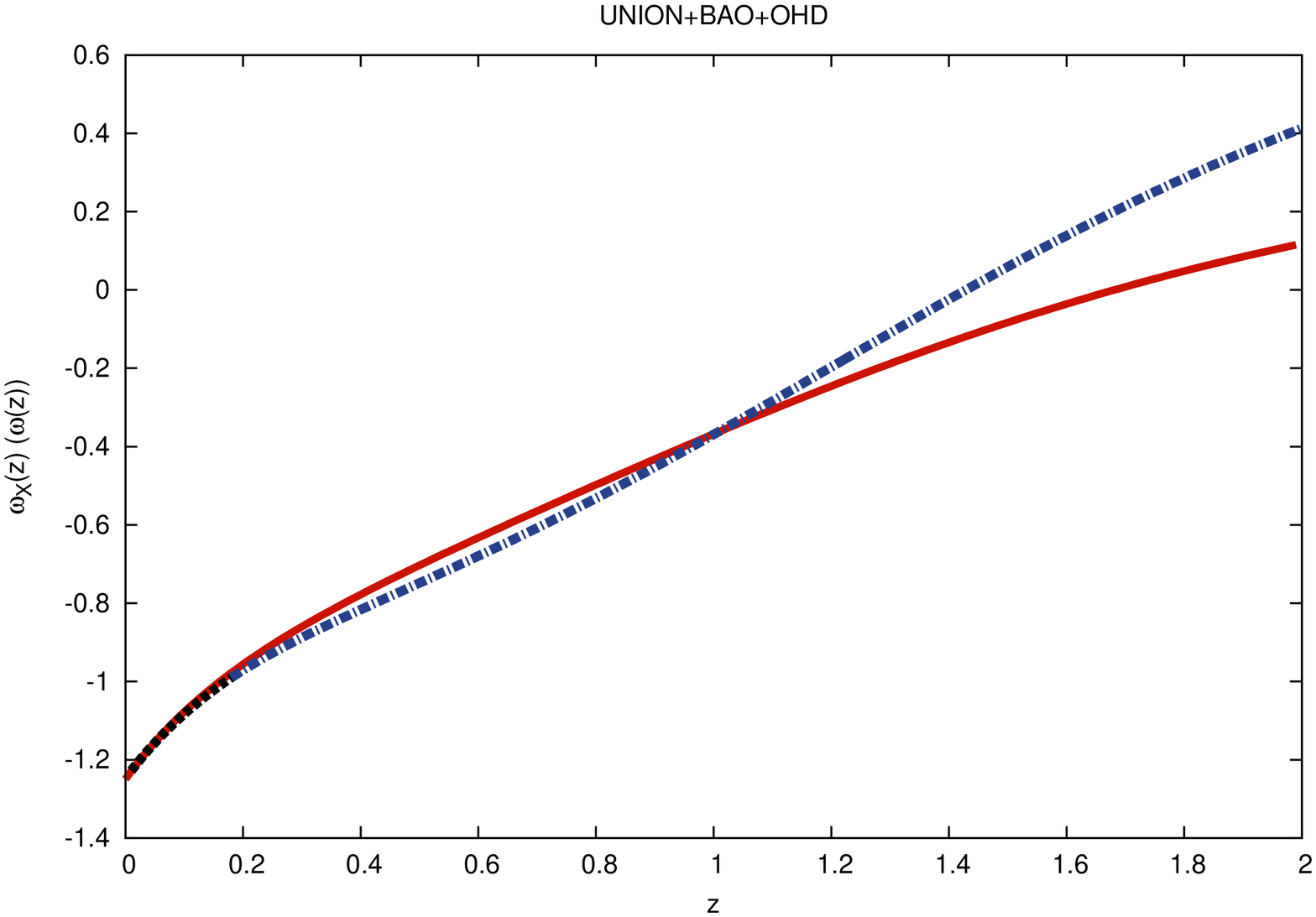}
\includegraphics[height=8cm,width=8cm,angle=-90]{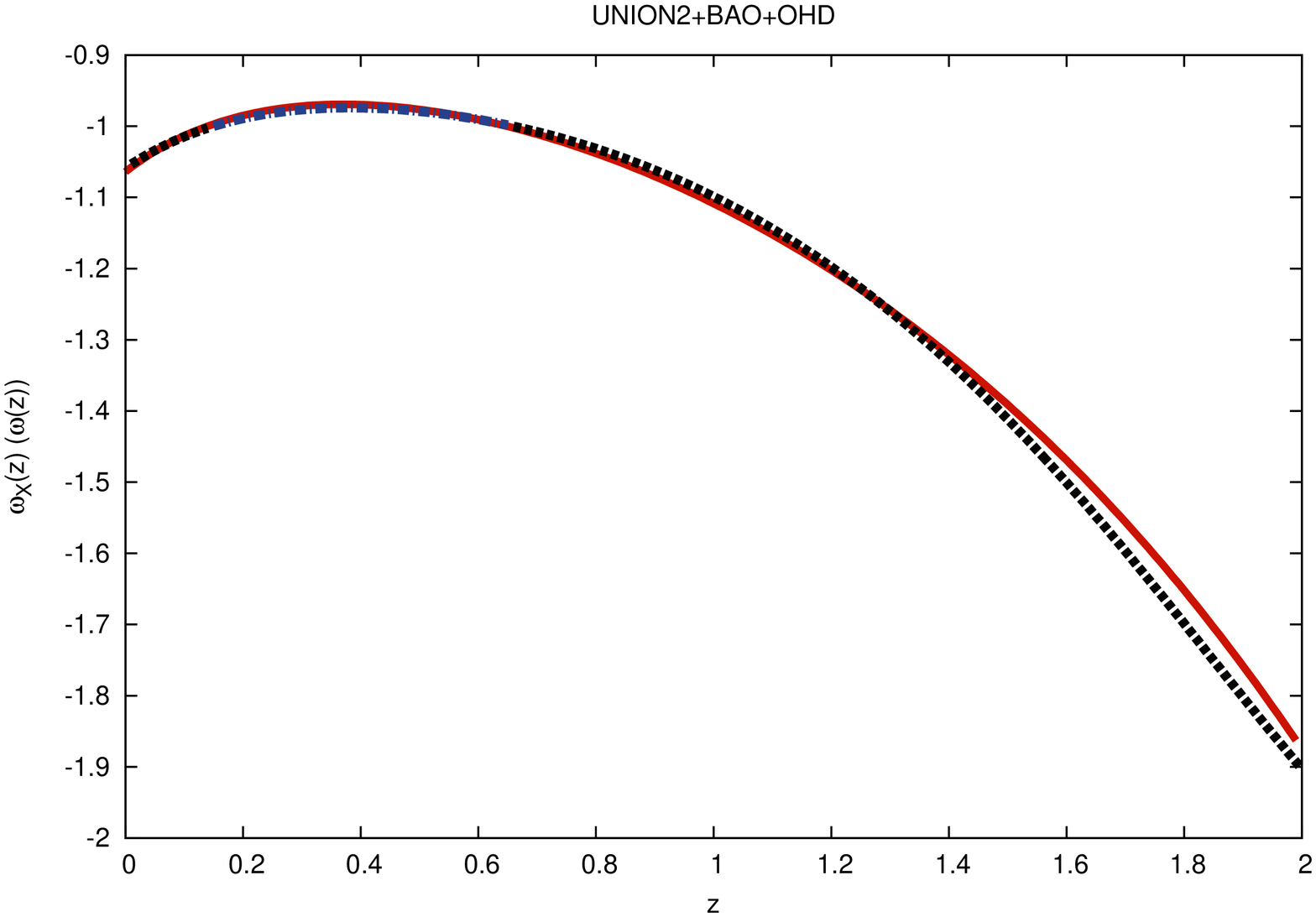}
\caption{Comparison of $\omega(z)$ vs $z$ and $\omega_X(z)$ vs $z$ for 
the 5 data sets considered (see text). $\omega(z)$ ($\omega_X(z)$) are plotted
along $y$ axis.}
\end{figure}

Using the formalism described in Sect. 4, the $\chi^2$ fit was performed 
\cite{taup}
for each of the five sets of SNe Ia data alongwith BAO and OHD data and
the best fit values of the parameters $a$, $b$ and $\Omega^0_m$
obtained for all the five sets \cite{taup}.
The best fit parameters for each set (alongwith BAO and OHD) are given in 
Table 1. From the obtained best fit values of the parameters the equation of 
state $\omega_X(z)$ is computed using Eq. (\ref{omegX1}) and 
Eq. (\ref{dlzparam}). Here we mention again, that no chosen 
form for $\omega_X(z)$ was adopted and $\omega_X(z)$ was obtained 
in a model independent way.
\begin{center}
\begin{tabular}{|cc|   cc|   cc|   cc|}
\hline\hline SNe Ia datasets && $a$ && $b$ &&
$\Omega_m^0$  &\\
(+BAO+OHD)&&&&&&& \\
\hline
HST+SNLS+ESSENCE     &&1.437&&0.550&&0.268&\\
(+BAO+OHD)&&&&&&& \\
(Data Set I)&&&&&&& \\ 
\hline
SALT2(+BAO+OHD) &&1.401&&0.542&&0.272&\\
(Data Set II)&&&&&&& \\
\hline
MLCS(+BAO+OHD)&&1.401&&0.653&&0.296&\\
(Data Set III)&&&&&&& \\
\hline
UNION(+BAO+OHD)&&1.635 &&0.699 &&0.268 & \\
(Data Set IV)&&&&&&& \\
\hline
UNION2(+BAO+OHD)&&1.289 &&0.458 &&0.272 & \\
(Data Set V)&&&&&&& \\
\hline
\end{tabular}
\end{center}
{\bf Table 1.} Best fit values from analysis of different data sets
\vskip 2mm

It is seen that except for Set I, for all other 
sets, $\omega_X(z)$ extends below the limit $\omega_X(z)=-1$ \cite{taup}.

As discussed earlier,     
the dark energy equation of state ($\omega(z)$) is also formulated  
from a quintom scalar field 
model in this work. This model is based on the fact that when 
$\omega(z) > -1$ its 
nature can be described by assuming a potential $V(\phi)$ for a scalar field 
$\phi$ responsible for the dark energy. While in the scenario when 
$\omega(z) < -1$ the dark energy is described by a potential $V(\sigma)$ 
for a phantom field $\sigma$. This is described in Sections 3.3 and 4.1.
Our purpose in this work is to compare dark energy equation of state 
($\omega(z)$) from
quintom scalar field model to that ($\omega_X(z)$) obtained from the data
and check ow well the present formalism for dark energy can explain 
the variation of $\omega_X(z)$.


From the the best fit values of the parameters tabulated in Table 1,
$d_L(z)$ and subsequently Hubble parameter $H(z)$ are computed.
As discussed in Sect. 3.3,
the fields $\phi(z)$ and $\sigma(z)$ can then be calculated
using Eqs. (\ref{expqu1},\ref{expph1}).
Consequently the chosen
forms of $V(\phi)$ (= $\exp[\lambda_1(1 + ((\phi-\phi_c)/M_{pl})^2)]$) 
and $V(\sigma)$ (= $\exp[\lambda_2(1 + ((\sigma-\sigma_0)/M_{pl})^2)]$) 
are calculated for 
the particular case when $\lambda_1 = \lambda_2 = 1$.  
The analyses of the five sets of data \cite{taup} indicate that for
Set II $-$ Set V, we need to consider quintom scalar field (since $\omega_X(z)$
goes below $-1$) while from data set I only the quintessence field suffices.

It is argued earlier (Sect. 4.1) that 
$\phi(z) = \phi(z_c)$, when $\omega_X(z) = -1$ and
$\phi(z)=0$ at other values of $z < z_c$ for which $\omega_X(z) < -1$.
Hence, $\phi(z_c)$ is used instead of $\phi_0$ in the entire calculation 
for the case $z > z_c$. On the other hand, $\sigma_0$, the value of 
$\sigma(z)$ at $z=0$,
remains undetermined and hence in our calculations,
we rescale $\sigma(z)$ as ($\sigma(z) - \sigma_0$) and show our results. 
 
We use the best fit values of the parameters given in Table 1 
for all the five sets of data and compute $\omega(z)$ formulated from 
quintom model following Eqs. (\ref{kequ},\ref{keph},\ref{quinomega}).
These are done for different values of $z$ in the range $0<z<1.76$.
For the present work, as stated earlier, $\lambda_1 = 1 = \lambda_2$ 
is chosen in the 
expression of the potential $V(\phi)$. 
They are then compared with corresponding $\omega_X(z)$ 
for the five data sets, obtained from the $\chi^2$ fit and using
Eqs (\ref{omegX1}, \ref{hth}) \cite{taup}. The results are shown in Fig. 1.

In Fig. 1 the red coloured (solid line) plots represent the variation
of $\omega_X(z)$ vs $z$ as obtained from the combined $\chi^2$ 
analyses of SNe Ia, BAO and OHD data \cite{taup}. The blue plots 
(dot-dashed line) in Fig. 1 are obtained from the calculation of
$\omega(z)$ with the present formalism. As described earlier, for the first plot
of Fig. 1 (corresponding to data set I), only quintessence scalar field 
describes the variation of $\omega(z)$ with the redshift $z$ since 
$\omega(z) > -1$, always, for this particular case. 
For the rest four plots the data analyses results show that   
$\omega_X(z) < -1$ for some redshift values. Hence for comparison 
with $\omega(z)$, the latter is calculated 
for quintom scalar field involving both $\phi(z)$ and $\sigma(z)$
for those four cases in Fig. 1 (corresponding to 
data Set II to Set V). In these four plots 
$\omega(z)$ from our calculations are given in bi-colour plots 
where the blue coloured dot-dashed line is obtained from the 
quintessence field
$\phi(z)$ ($\omega(z) \geq -1$) and calculations with the 
phantom field ($\sigma(z)$) are represented
by only black dashed line ($\omega(z) < -1$) of the plot. 

From Fig. 1 one readily sees that the variations of the 
dark energy equation of states
calculated using the adopted forms of $V(\phi)$ and $V(\sigma)$ 
in the present quintom scalar field formalism, are continuous.  
There are no discontinuities even when $\omega(z)$ varies from  
$\omega(z) > -1$ regime to the region where $\omega(z) < -1$. More importantly,
it is also evident from Fig. 1 that the dark energy equation 
of states $\omega(z)$ calculated from the proposed quintom scalar field theory 
in the present work are in good agreements with $\omega_X(z)$ for 
at least three 
cases of data sets (namely Sets II, III and V) including the 
most recent UNION2 data set considered here. These results indicate 
that the proposed quintom scalar field formalism can be a viable model
for explaining the varying dark energy.

 

\thispagestyle{empty}

\end{document}